\documentclass[a4paper,UKenglish,cleveref, autoref, thm-restate]{lipics-v2021}

\usepackage{subcaption}
\usepackage{xcolor}
\usepackage{tikz}
\usetikzlibrary{arrows.meta, positioning, fit, backgrounds, shapes.misc, calc}

\title{An Agentic Approach Towards Replication Package Quality Evaluation}

\titlerunning{An Agentic Approach Towards Replication Package Quality Evaluation}

\author{Maximilian Alexander Amougou Mbida}{Technical University of Munich, Germany \and \url{https://github.com/CallMeSwarley}}{maxi.amougou@tum.de}{https://orcid.org/0009-0003-3831-1007}{}

\author{Florian Angermeir}{fortiss, Germany \and Blekinge Institute of Technology, Sweden \and \url{https://angermeir.me} }{angermeir@fortiss.org}{https://orcid.org/0000-0001-7903-8236}{}

\authorrunning{M. Amougou et al.}

\Copyright{Maximilian Alexander Amougou Mbida and Florian Angermeir}



\nolinenumbers

\begin{document}

\maketitle

\begin{abstract}
Reproducibility in empirical software engineering relies on complete, accessible, and reusable research artifacts, yet artifact evaluation remains largely manual and difficult to scale. 
This emerging results paper explores an agentic approach for assessing replication package quality by translating open-science guidelines into machine-verifiable criteria. 
We consolidate 380 requirements from 34 sources into 51 reproducibility criteria, of which 31 are operationalized for automated artifact-based evaluation. 
Based on these criteria, we implement a multi-agent prototype that inspects replication packages and produces evidence-grounded improvement reports. 
A preliminary evaluation on five replication packages shows high inter-run consistency of 91.4\% and 75.4\% correctness, through micro-averaged agreement with a manual baseline. 
The agent performs best on structural criteria such as code, environment, and artifact availability, but struggles with qualitative or mixed-method studies. 
A pilot survey with seven software engineering researchers indicates well perceived usefulness and adoption potential, while revealing cognitive load in the human-in-the-loop planning step. 
Overall, within these small samples, the results indicate that agentic research artifact evaluation has the potential to support authors and reviewers by automating selected routine checks.
\end{abstract}

\section{Introduction}
\label{sec:introduction}
Reproducibility is a fundamental prerequisite for cumulative scientific progress, enabling independent verification of findings and fostering confidence in published research \cite{national2019reproducibility, da2014replicationcornerstone}. In large parts of empirical software engineering, reproducibility depends not only on textual reporting, but also on the availability and quality of accompanying research artifacts such as source code, datasets, execution scripts, configuration files, and experimental environments \cite{winter2022retrospective}. To enhance transparent research practices, conferences and publishers have introduced artifact-evaluation tracks, open science guidelines, and reporting standards \cite{ralph2020empirical, baltes2025guidelinesempiricalstudiessoftware}.
Despite these efforts, research artifacts frequently suffer from a plethora of issues, including incomplete source code \cite{liu_chao, GONZALEZBARAHONA2023107318}, missing dependencies \cite{trisovic2022large, HASSAN2025112327}, and fragmented documentation \cite{gonzalez2012reproducibility}. Consequently, verifying whether these packages are sufficiently complete, accessible, and reusable remains a manual, time-consuming process that scales poorly and is subject to interpretation~\cite{wu2026agentbasedsoftwareartifactevaluation, winter2022retrospective, acm_badging, siegel2024corebenchfosteringcredibilitypublished}.

While existing guidelines outline what researchers should report, they provide limited support for automatically assessing the quality of heterogeneous artifact types against predefined criteria representing open science best practices. As a result, artifact evaluation continues to impose substantial effort on reviewers, slows down review processes, and creates barriers for large-scale reproducibility assessments. This challenge is particularly pronounced in contemporary software engineering research, where replication packages lack standardization and often bundle a mix of source code, raw datasets, and configuration files across various directories and formats.

In this paper, we explore whether an agentic approach can support replication package quality evaluation by operationalizing open-science guidelines into machine-actionable verification procedures. We propose an agent-based approach that autonomously retrieves, structures, and evaluates research artifacts against reproducibility criteria, enabling a scalable assessment of research artifacts quality as a proxy to the potential for reproducibility. We target computational, code-centric replication packages containing inspectable artifacts such as source code, datasets, and configuration files. We do not aim to assess the context-dependent rigor of qualitative or theoretical artifacts. Conceptually, our approach can act as a ``replication package linter'': it flags evidence-backed quality issues before formal artifact review rather than establishing that a study's results are reproducible. We hope that our contribution in the long-term can support authors in improving their research artifact quality even before artifact review and to consequently reduce review overhead in artifact review tracks. To investigate the feasibility of this idea, we implemented a prototype based on a multi-agent architecture and conducted a pilot survey. Our preliminary evaluation indicates promising results. The approach achieves an inter-run consistency of 91.4\% and a 75.4\% micro-averaged agreement with manual baseline. While still preliminary, these results strengthen our confidence in the potential and long-term value of the proposed approach.

The main contributions of this paper are (1) an operationalization of open-science guidelines into a set of machine-verifiable criteria for automated artifact assessment, (2) a novel agentic approach for assessing the reproducibility potential of software engineering research artifacts, and (3) preliminary empirical evidence from a pilot survey that demonstrates the potential and limitations of our approach.

\section{Related Work}
\label{sec:background}
We structure existing related work into challenges with manual and static artifact evaluation, and autonomous reproduction systems.

\noindent \textbf{Challenges and Automated Artifact Evaluation.} 
Achieving sustainable reproducibility is hindered by the post-publication fragility of long-term archiving \cite{winter2022retrospective} and a pre-publication reliance on manual human labor during peer review. This constraint leads to inconsistent evaluation depth and limits scalability, driving the necessity for automated review assistance. While earlier automation focuses on static analysis of manuscripts, such as Metacheck \cite{debruine2025metacheck}, such solutions can be limited by the lacking of semantic analysis depth of research artefacts. Our approach resolves this through fully automated, end-to-end evaluation of both the artifact content and the structural alignment.

\noindent \textbf{Agentic Reproducibility Benchmarks and Frameworks.}  Recent benchmarks quantify LLM capabilities in scientific reproduction, typically reporting low success rates (20\% to 40\%) that highlight domain complexity. Evaluation paradigms target either execution and consistency checks (such as dependency installation in CORE-Bench \cite{siegel2024corebenchfosteringcredibilitypublished} and paper-to-claim alignment in REPRO-BENCH \cite{hu2025repro}), or the harder task of generating reproduction code from scratch, as evaluated by PaperBench \cite{starace2025paperbenchevaluatingaisability} and LMR-BENCH \cite{yan2025lmr}.
To tackle these complexities, several architectures employ specialized reasoning pipelines. For instance, AutoReproduce \cite{zhao2025autoreproducereproducebench} and ResearchCodeAgent \cite{gandhi2025research} leverage paper lineages and execution feedback to iteratively refine scripts, while SciRep/SciConv \cite{costa2025framework, costa2025talk} automate Docker-based setups. Closest to our work, ArtifactCopilot \cite{wu2026agentbasedsoftwareartifactevaluation} maps unstructured documentation to dependency graphs for end-to-end ACM badging \cite{acm_badging}.
However, most of the agentic approaches assume well-structured repositories, complete documentation, or explicit instructions. In contrast, our framework focuses on evaluating the \emph{reproducibility potential} of heterogeneous, inconsistently structured research artifacts where documentation may be partial with the goal of enhancing artifact quality.

\section{Methodology}
\label{sec:methodology}
To guide the design, implementation, and evaluation of our approach, we organized our methodology around three central research questions:
\begin{itemize}
    \item[]\textbf{RQ1:} How can an agentic approach assess the potential for reproducibility given a research artifact set?
    \item[]\textbf{RQ2:} What are the potential and limitations of an agentic approach to assess the quality of a replication package?
    \item[]\textbf{RQ3:} How useful is the agentic approach for stakeholders assessing replication package quality?
\end{itemize}

\paragraph*{Reproducibility Criteria}
To establish robust reproducibility criteria, we analyzed official guidelines and reviewer resources from top-tier (CORE A/A* \cite{core_portal2025}) computer science conferences featuring artifact evaluation tracks or explicit checklists (e.g., ICSE, OSDI, ICML, RE). This was supplemented by open science policies from major publishers, leading empirical software engineering journals (EMSE, IST, IEEE Access \cite{scimago_sjr2025}), and community-driven best practices.

\noindent \textbf{Extraction Process.} We categorized the source documents into four groups: \textbf{(1)} conferences with artifact badges, \textbf{(2)} checklist-only venues, \textbf{(3)} journals and publishers, and \textbf{(4)} general guidelines. Verbatim criteria were extracted into a centralized repository, semantically merged within each category, and then consolidated into a unified, hierarchical reproducibility criteria knowledge base. We assigned each criterion a unique identifier to maintain full traceability to its original source, and organized them into hierarchical topics and subtopics. Finally, because raw textual descriptions were insufficient for automated reasoning, we enriched the criteria with structured attributes to enable systematic evaluation by our agent.

For example, the raw requirements ``A ReadMe file that includes a table of results accompanied by precise command to run to produce those results'' (\texttt{fairer\_34}) and ``make it easy to reproduce all of the data for your paper, ideally with a single batch script or mouse click'' (\texttt{aec\_16}) were first normalized into requirements for step-by-step instructions and exact commands. These and related requirements were then merged into the criterion ``Provide detailed step-by-step instructions (e.g., in a README) to reproduce all results with exact commands.'' We enriched this criterion by specifying the required input as a README, installation or usage documentation, and example commands. The detection strategy checks for sequential instructions and executable command examples and the decision boundary requires a README or installation file containing commands for every experimental step. The database retains links from the merged criterion to all source requirements.

\noindent \textbf{Enrichment.} Raw textual criteria are often high-level and ambiguous, leading to high evaluation variance that is further exacerbated by LLM non-determinism. To ensure consistent assessment trajectories, we manually enriched each criterion with four structured auxiliary fields: 
\textbf{(1)} \texttt{additional\_information} (contextual scope and definitions), 
\textbf{(2)} \texttt{input\_needed} (abstract evidence types to seek), 
\textbf{(3)} \texttt{detection\_strategy} (initial evaluation heuristics), and 
\textbf{(4)} \texttt{success\_failure\_conditions} (explicit decision boundaries). 
We instantiated these specifications in an SQLite knowledge base that preserves relationships and source traceability, and supports relevance-based retrieval for the agent via the full-text search.

\paragraph*{Agentic System Design}

\begin{figure}
    \centering
    \makebox[\textwidth]{\resizebox{\textwidth}{!}{\begin{tikzpicture}[
    font=\scriptsize,
    >={Stealth[length=2mm]},
    node distance=4mm,
    proc/.style={rectangle, rounded corners=2pt, draw, line width=0.5pt,
                 fill=blue!8, align=center, inner sep=3pt, minimum height=6mm},
    term/.style={rounded rectangle, draw, line width=0.5pt,
                 fill=black!10, align=center, inner sep=3pt, minimum height=6mm},
    hitl/.style={rectangle, rounded corners=2pt, draw, line width=0.5pt,
                 fill=orange!20, align=center, inner sep=3pt, minimum height=6mm},
    note/.style={rectangle, draw, dashed, line width=0.4pt, fill=yellow!18,
                 align=left, inner sep=3pt, font=\scriptsize\itshape},
    flow/.style={->, line width=0.7pt},
    fail/.style={->, line width=0.7pt, densely dotted, draw=red!55!black},
    notelink/.style={dashed, line width=0.4pt, draw=black!55},
    elab/.style={font=\scriptsize, fill=white, inner sep=1pt},
    phasebox/.style={draw, dashed, rounded corners=4pt, line width=0.4pt, inner sep=4mm},
    ptitle/.style={font=\footnotesize\bfseries, anchor=south west, inner sep=1pt}
]

\node[term] (p1_s)   at (0,0)      {\texttt{start}};
\node[proc] (p1_c)   at (2.2,0)    {\texttt{convert\_pdf}};
\node[proc] (p1_sum) at (4.7,0.8)  {\texttt{summarize\_paper}};
\node[proc] (p1_v)   at (4.7,-0.8) {\texttt{vectorize\_paper}};

\node[proc] (p2_g)  at (7.7,0.8)    {\texttt{get\_artifact}};
\node[proc] (p2_rm) at (11.3,0.8)   {\texttt{get\_artifact\_readme}};
\node[proc] (p2_v)  at (15.3,0.8)   {\texttt{vectorize\_artifact}};
\node[hitl] (h_art) at (7.7,-0.8)   {\texttt{hitl\_artifact}};
\node[hitl] (h_rdm) at (11.3,-0.8)  {\texttt{hitl\_readme}};

\node[proc] (p3_w) at (0,-5.6)    {\texttt{plan\_worker}};
\node[proc] (p3_s) at (2.7,-5.6)  {\texttt{plan\_synthesizer}};
\node[hitl] (p3_a) at (6.5,-5.6)  {\texttt{hitl\_approval\_node}};
\node[proc] (p3_r) at (6.5,-7.1)  {\texttt{planning\_router}};
\node[proc] (p3_m) at (3.0,-7.1)  {\texttt{modify\_plan}};

\node[proc] (p4_e)   at (11.0,-4.6)  {\texttt{execute\_plan}};
\node[note, text width=2.9cm] (p4_n) at (14.1,-4.6)
    {Internal: \texttt{orchestrator\_worker} synchronously invokes sub-agents};

\node[proc] (p5_g)   at (11.0,-7.4) {\texttt{generate\_report}};
\node[term] (p5_end) at (13.4,-7.4) {\texttt{end}};

\draw[flow] (p1_s) -- (p1_c);
\draw[flow] (p1_c) -- (p1_sum);
\draw[flow] (p1_c) -- (p1_v);

\draw[flow] (p2_g)  -- (p2_rm) node[elab,midway,above]{Success};
\draw[flow] (p2_rm) -- (p2_v)  node[elab,midway,above]{Success};
\draw[flow] (p2_g)  to[bend right=25] node[elab,left,xshift=4mm]{Not Found} (h_art);
\draw[flow] (h_art) to[bend right=25] node[elab,right]{URL} (p2_g);
\draw[flow] (p2_rm) to[bend right=25] node[elab,left]{Not Found} (h_rdm);
\draw[flow] (h_rdm) to[bend right=25] node[elab,right]{Path} (p2_rm);

\draw[flow] (p3_w) -- (p3_s);
\draw[flow] (p3_s) -- (p3_a) node[elab,midway,above]{Success};
\draw[flow] (p3_a) to[bend right=26] node[elab,left]{Feedback} (p3_r);
\draw[flow] (p3_r) to[bend right=26] node[elab,right]{No Modification} (p3_a);
\draw[flow] (p3_r) -- (p3_m) node[elab,midway,above]{Modify};
\draw[flow] (p3_m) -- (p3_a) node[elab,midway,sloped,above]{Modified Plan};

\draw[flow] (p4_e) -- (p5_g);
\draw[flow] (p5_g) -- (p5_end);
\draw[notelink] (p4_e) -- (p4_n);

\draw[flow] (p1_sum) -- (p2_g);
\draw[flow] (p1_v)   -- (p2_g);
\draw[flow, rounded corners=8pt] (p2_v.south) -- ++(0,-3.0) -| (p3_w.north)
    node[elab, pos=0.7, above]{Success};
\draw[flow] (p3_a) -- (p4_e) node[elab,midway,above]{Approve};

\draw[fail] (p2_v.east)   -- (17.0,0.8) -- (17.0,-6.8) -| ($(p5_g.north)+(0.7,0)$);
\draw[fail, -] (p2_g.east)  -- ++(0.3,0) -- ++(0,-2.6) -| (17.0,-2.6);
\draw[fail, -] (p2_rm.east) -- ++(0.3,0) -- ++(0,-2.6) -| (17.0,-2.6);
\node[elab, text=red!55!black] at (17,-4.0) {Fail};
\draw[fail] ($(p3_s.south)+(-0.9,0)$) |- (8.8,-8.0) |- (p5_g.west)
    node[elab, text=red!55!black, pos=0.3, below]{Fail};

\begin{pgfonlayer}{background}
    \node[phasebox, fill=blue!4,   fit=(p1_s)(p1_c)(p1_sum)(p1_v)] (boxP1) {};
    \node[phasebox, fill=cyan!4,   fit=(p2_g)(p2_rm)(p2_v)(h_art)(h_rdm)] (boxP2) {};
    \node[phasebox, fill=violet!4, fit=(p3_w)(p3_s)(p3_a)(p3_r)(p3_m)] (boxP3) {};
    \node[phasebox, fill=orange!4, fit=(p4_e)(p4_n)] (boxP4) {};
    \node[phasebox, fill=green!4,  fit=(p5_g)(p5_end)] (boxP5) {};
\end{pgfonlayer}

\node[ptitle] at (boxP1.north west) {Phase 1: Document Ingestion};
\node[ptitle] at (boxP2.north west) {Phase 2: Artifact Retrieval};
\node[ptitle] at (boxP3.north west) {Phase 3: Plan Generation};
\node[ptitle] at (boxP4.north west) {Phase 4: Subtopic Evaluation};
\node[ptitle] at (boxP5.north west) {Phase 5: Report Compilation};

\end{tikzpicture}}}
    \caption{Overview of the five-phase agentic evaluation pipeline. Solid lines indicate standard execution and progression pathways, while dotted lines represent early-exit failure or abort routing back to the report generation phase.}
    \label{fig:agent_workflow}
\end{figure}
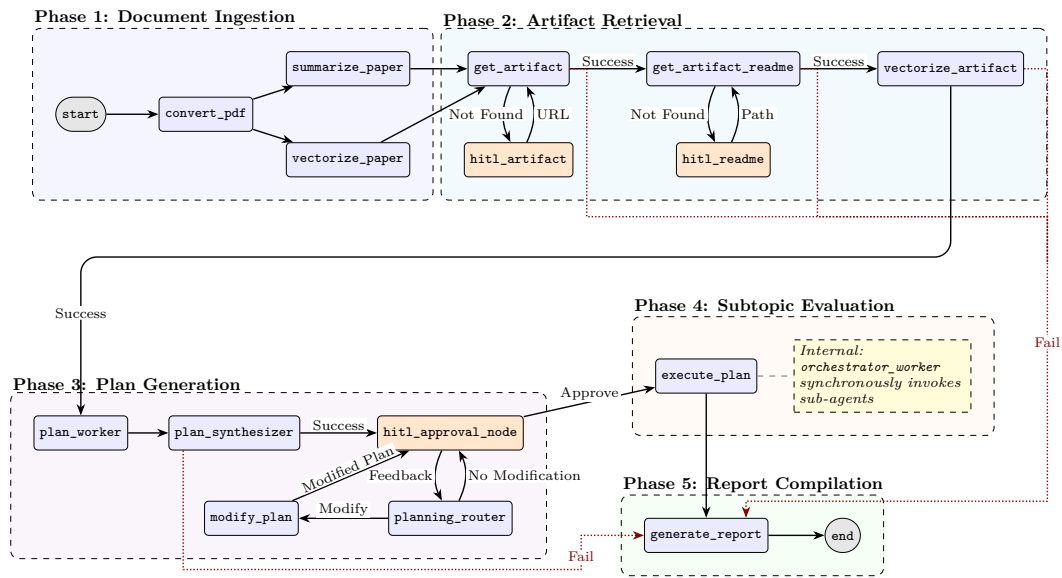

The approach was implemented as a stateful, graph-based workflow. As illustrated in \autoref{fig:agent_workflow}, the architecture consists of a five-phase pipeline: (1) ingestion and embedding, (2) artifact retrieval and analysis, (3) plan generation, (4) subtopic evaluation, and (5) report compilation. The workflow leverages LangGraph's \texttt{Command} API for dynamic routing and an \texttt{interrupt} mechanism for synchronous Human-in-the-Loop (HITL) interactions, persisting agent state to ensure conversation continuity.

\noindent \textbf{Phase 1: Ingestion and Embedding.} The input paper is converted from PDF to markdown to extract key metadata (title, abstract, artifact details) and a global summary. Concurrently, the text is chunked and embedded into a vector database for retrieval-augmented generation (RAG) throughout execution.

\noindent \textbf{Phase 2: Artifact Retrieval and Analysis.} A sub-agent extracts artifact references (e.g., GitHub, Zenodo, OSF) from the paper summary. If automatic retrieval or entrypoint (e.g., README) identification fails, a HITL interrupt requests user guidance, e.g., the correct artifact URL, an upload, or the explicit entrypoint path. If no entry file exists, the agent logs a failure, and skips to an error report. The retrieved README is summarized to ground planning, a multi-stage filter then prunes the vector index, combining fast rule-based exclusion of  irrelevant files (e.g., binaries, dependencies, logs) with an LLM-based classifier for extensionless or ambiguous files. The remaining relevant files are chunked and embedded for hybrid retrieval in the next phase.

\noindent \textbf{Phase 3: Evaluation Plan Generation.} Worker sub-agents analyze the retrieved context and concurrently generate per-criteria inspection steps for the user-selected subtopics, synthesized into a unified, hierarchical strategy conforming to the knowledge base layout. A HITL loop then lets users iteratively refine this strategy via natural language feedback, suspending execution until explicit user approval.

\noindent \textbf{Phase 4: Parallel Subtopic Evaluation.} An orchestrator dispatches sub-agents to evaluate each subtopic concurrently within isolated reasoning contexts over a shared, read-only filesystem, gathering evidence via pattern search and hybrid retrieval (lexical and semantic). Evaluation demands evidence-based reasoning with explicit citations (e.g., file paths, code snippets), defaulting to a \emph{fail} status if required evidence is missing. Findings are then synthesized into subtopic- and topic-level summaries and a final structured output.

\noindent \textbf{Phase 5: Report Compilation and Tracing.} The structured output is transformed into a hierarchical markdown report aggregating evaluation metadata, paper summaries, and detailed criterion assessments linked to their guidelines. For transparency and reproducibility, it logs the complete execution plan and embeds Langfuse trace IDs.

\noindent \textbf{Frontend.} A web-based prototype exposes four views mirroring the agent's workflow: configuration, planning (the HITL loop), the execution (real-time reasoning and tool invocations), and  the report dashboard.

\noindent \textbf{Agent Models and Configurations}
The approach uses five agents: artifact retrieval, plan, orchestrator, executor, and reporting.
For cost reasons, the retrieval and the executor agent use gpt-5-nano-2025-08-07, the plan agent and orchestrator agents use gpt-5-mini-2025-08-07. The reporting agent uses gpt-5.1-2025-11-13. We set the temperature for all agents to 0. The full agent setup is published in the replication package \cite{online_material}. 

\paragraph*{Evaluation}

\noindent \textbf{Reliability and Consistency.} 
To establish the technical reliability of the automated evaluation, we measured the approach's \textit{inter-run consistency} and \textit{agreement with a manual baseline} on a small, purposive sample of five replication packages. We selected packages from publications familiar to the authors to support detailed manual assessment. This convenience sample is not representative of replication packages in software engineering, and familiarity may have made the manual assessment easier or more favorable than an assessment of unfamiliar artifacts. Consequently, the resulting scores are exploratory and should not be interpreted as population-level performance estimates.

To evaluate inter-run consistency, we executed the agentic pipeline 10 independent times per package. Stability across identical inputs was evaluated for the execution plan and the evaluation report on a per-element basis for each artifact. We evaluated those elements in two ways. Where possible we used deterministic classification (e.g., Pass/Fail for a subtopic). Here stability was calculated using a categorical overlap score that aggregates consensus via a bottom-up hierarchical macro-average. Where deterministic classification was impossible (e.g., for the improvement suggestions) we employed an LLM-as-Judge (gpt-5.4-mini-2026-03-17) to evaluate the semantic consistency across randomly paired runs. This judge analyzed the alignment of the agent's generated outputs on a 1--4 rubric (1 = Not Consistent, 4 = Fully Consistent). We verified the judge's reliability against a manually created sample, achieving approximately 85\% accuracy in score assignment. All scripts, judge prompts, and raw data logs are available in the online material \cite{online_material}. 

Second, to quantify the agents agreement with the manual baseline, the authors manually assessed the same five replication packages using the operationalized criteria to establish a structured ground-truth dataset. 
The agent's classifications were benchmarked against this manual baseline. To capture granular performance across distinct criteria states (Pass, Fail, Not Applicable), we computed two alignment metrics: a macro-accuracy (to enforce strict schema matching and weights all hierarchical topics equally) and a standard micro-accuracy (to evaluate a flattened intersection of shared criteria, weighting every individual criteria equally). 
These deterministic metrics were supplemented by the aforementioned LLM-as-Judge, which similarly scored the semantic alignment of the agent's generated qualitative reports against the human baseline. These measures quantify correspondence with an author-created reference under our operationalized criteria, they do not establish objective correctness, or completeness of the recommendations.

\noindent \textbf{Perceived Usefulness and Usability.} To assess the practical value of the generated reports, we conducted a survey with active software engineering researchers, ranging from PhD candidates to professors. Participants were recruited via direct email invitations. Prior to interacting with the approach, participants were provided with instructions and context. They were briefed on the prototypes' objective, provided with credentials for the hosted instance of the prototype, and asked to evaluate one of their own recent research artifacts. Crucially, to ensure data privacy, participants were instructed to avoid uploading confidential or unpublished materials due to the prototypes reliance on commercial LLMs.

Following this setup, participants used the prototype and evaluated their chosen artifact by interacting with the agent's evaluation plan and reviewing the final report. Participants then completed a structured questionnaire, which was informed by the Technology Acceptance Model (TAM) \cite{TAM}, comprising five-point Likert-scale items and open-ended questions. This allowed us to evaluate the prototypes perceived usefulness and quality (assessing the accuracy, actionability, and detail of the feedback) alongside its usability and adoption potential (ease of use, interaction design, and the likelihood of integration into regular workflows). The survey instrument and anonymized responses are published in the online material \cite{online_material}.

\section{Emerging Results}
\label{sec:results}
This section presents our findings structured around our three research questions: the formulation of the reproducibility criteria (RQ1), an analysis of the agent's strengths and limitations (RQ2), and an empirical assessment of the approach’s perceived usefulness to stakeholders (RQ3).

\paragraph*{RQ1: Reproducibility Criteria}
We distill an initial pool of 380 raw requirements from 34 sources into a unified set of 51 reproducibility criteria. Because some requirements demand paper-text analysis rather than artifact inspection, we distinguish between the full set and a subset of 31 operationalized criteria that are fully amenable to automated, artifact-based evaluation.
The extracted criteria are organized into a hierarchical structure comprising five high-level topics and fourteen subtopics. In the list below, the numbers in parentheses indicate the total count of individual criteria assigned to each respective topic and subtopic: \\
\noindent \textbf{Documentation (15):} Covers the core information required to comprehend and reproduce the research. Subtopics: \emph{Theoretical \& Mathematical} (7), \emph{Data} (5), and \emph{Experimental \& Methodological} (3).

\noindent\textbf{Data and Artifact Availability (16):} Focuses on ensuring artifacts remain accessible, reusable, and secure over time. Subtopics: \emph{Archiving \& Accessibility} (6), \emph{Data Integrity \& Privacy} (4), \emph{Licensing \& Citation} (2), and \emph{Reproducibility Support} (4).

\noindent\textbf{Code and Environment (7):} Addresses the executability of the software artifact. Subtopics: \emph{Automatization} (1), \emph{Code Availability \& Management} (3), \emph{Integration \& Documentation} (1), and \emph{Reproducibility \& Infrastructure} (2).

\noindent\textbf{Experimental Rigor (7):} Captures transparency and reliability of empirical setups. Subtopics: \emph{Experimental Reliability \& Statistical Reporting} (2), \emph{Hyperparameter Transparency} (3), and \emph{Model \& Randomness Control} (2).

\noindent\textbf{Ethical and Governance (6):} Encompasses ethical design and reporting considerations. Subtopics: \emph{Human-Subject Research} (2) and \emph{Responsible Use of Models \& Code} (4).

\paragraph*{RQ2: Strengths and Limitations}
To answer RQ2, we analyze the potential and limitations of using the agentic approach to assess the quality of replication packages.

\noindent \textbf{Strengths.}
The approach shows strong potential in automating the foundational, labor-intensive steps, particularly for computational and code-centric research. Its retrieval pipeline robustly ingests diverse sources, autonomously cloning GitHub repositories, fetching Zenodo archives, and parsing direct URLs or manual uploads.

\noindent \textbf{Inter-Run Consistency (Stability).}
The agent demonstrated high determinism across repeated executions.  Using a hierarchical macro-averaging metric that follows our predefined knowledge base schema (i.e., calculating categorical overlap exactly across the established topics, subtopics, and criteria), the approach achieves an overall consistency score of 91.4\% across all runs. Stability was particularly strong in structured domains, such as \emph{Data and Artifact Availability} (94.8\%) and \emph{Ethical and Governance} (94.0\%).

When evaluating via the LLM-as-Judge semantic rubric (scored 1--4), the agent's high-level qualitative outputs proved similarly stable. Applying this validated judge, the agent's natural language reasoning (3.36/4.0) and replication improvement suggestions (3.23/4.0) consistently achieve a "Mostly Consistent" rating across runs. Interestingly, the lowest stability is observed in exact evidence path extraction (2.64/4.0), indicating that while the agent's final reasoning and evaluations are deterministic, the localized text snippets it cites to justify those decisions can vary stochastically between runs.

\noindent \textbf{Limitations and Failure Cases.} 
Despite its strengths, testing revealed several limitations and failure cases driven by the tension between rigid automated evaluation and heterogeneous scientific reporting:

\begin{itemize}
    \item[] \textbf{Rigidity and Prompt Engineering Overhead:} To preserve internal consistency, the success and failure boundaries for criteria must be defined very strictly, which reduces flexibility. For example, evaluating whether data was appropriately anonymized initially resulted in inconsistent penalties for purely technical datasets. Preventing these false negatives required engineering the criteria's applicability conditions to explicitly distinguish between human-subject data and inherently synthetic data (e.g., regex syntax or physics simulations).
    
    \item[] \textbf{Contextual Blind Spots (The ``If Applicable'' Problem):} The approach struggles with conditional criteria. For instance, criteria requiring authors to provide pre-trained model weights or parameter counts caused false failures when the agent evaluated replication studies. Lacking the semantic awareness that the authors were re-running previous code rather than training a novel model, the agent incorrectly demanded non-existent artifacts.
    
    \item[] \textbf{Extraction and Attribution Ambiguity:} Artifact links frequently hide in footnotes, unstructured data availability sections, or inline citations, and are inconsistently labeled (e.g., ``online material'' versus ``replication package''). This occasionally causes the agent to misclassify a supplementary dataset as a primary code repository.
    
    \item[] \textbf{Preprocessing and Pipeline Fragility:} Edge cases in formatting can quickly exceed the context window size and processing constraints. During evaluation, oversized or malformed inputs (e.g., data-heavy PDFs, image-laden Jupyter Notebooks, and deeply nested archives) triggered an explosion of text chunks, system timeouts, or an empty workspace with subsequent execution faults. As a preliminary mitigation, we instruct users to provide flattened, non-nested archives prior to uploading their artifacts.
\end{itemize}

\paragraph*{RQ3: Usefulness for Stakeholders}
Since a report is only useful if its judgments are correct, we first benchmark the agent against a manual baseline, then survey researchers on perceived usefulness.

\noindent \textbf{Agreement with Manual Baseline.}
When comparing against human-authored baselines, the agent achieved a micro-accuracy of 75.4\% and a macro-average of 68.2\%.  Breaking down the classification metrics reveals performance differences. The agent is reliable at identifying compliant practices, achieving 86.4\% precision and 81.5\% recall ($F1 = 0.83$) when classifying criteria as ``Passed'' (True) in the final report. Conversely, when flagging non-compliant (``False'') criteria, it exhibited high recall (85.8\%) but lower precision (56.7\%), suggesting a slight bias toward strictness, which represents our prompting strategy, instructing the agent to default to false if the answer is not clear.

\begin{figure}[htbp]
    \centering
    \begin{subfigure}[t]{0.49\linewidth}
        \centering
        \includegraphics[width=\linewidth]{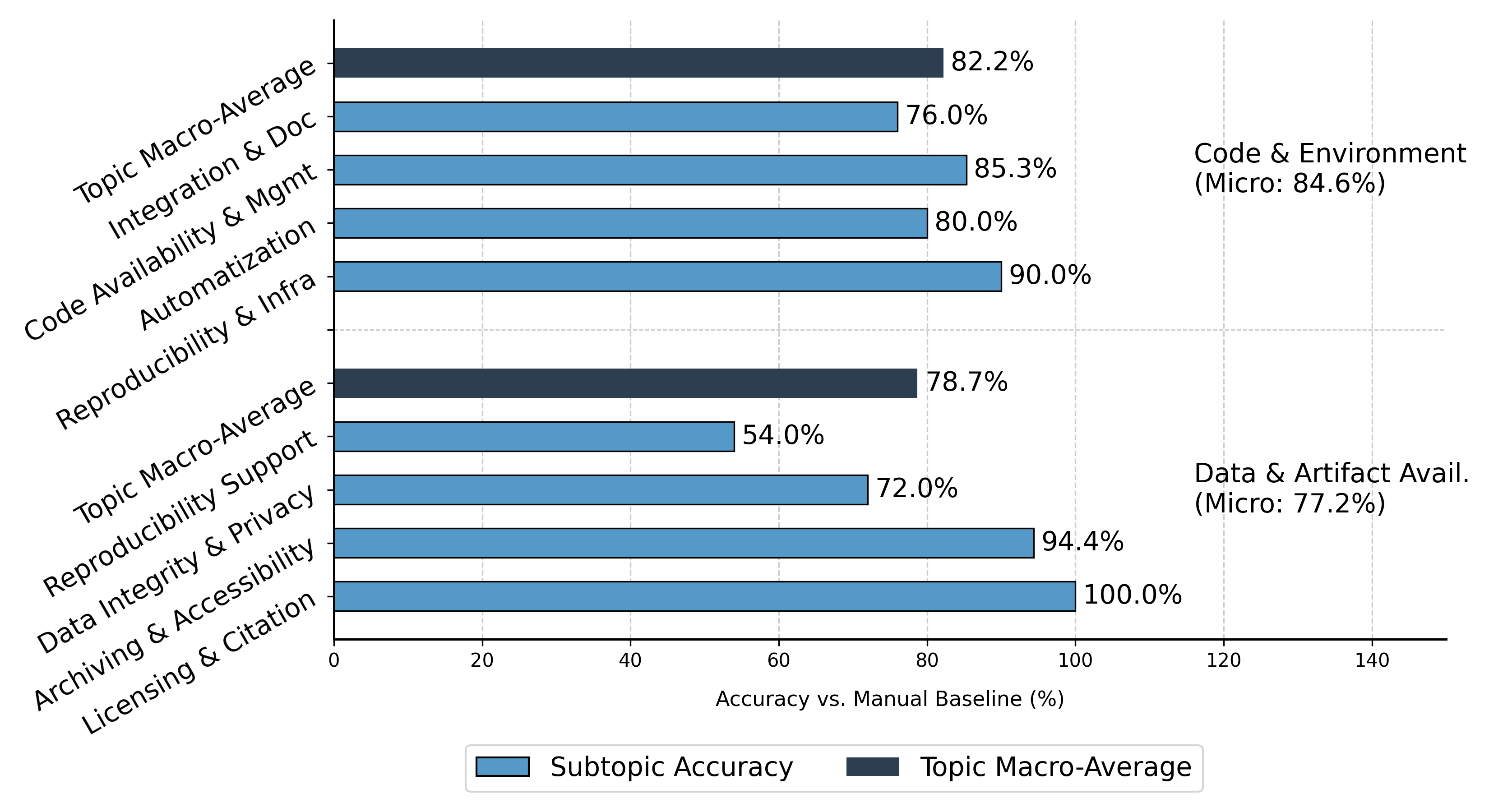}
    \end{subfigure}
    \hfill
    \begin{subfigure}[t]{0.49\linewidth}
        \centering
        \includegraphics[width=\linewidth]{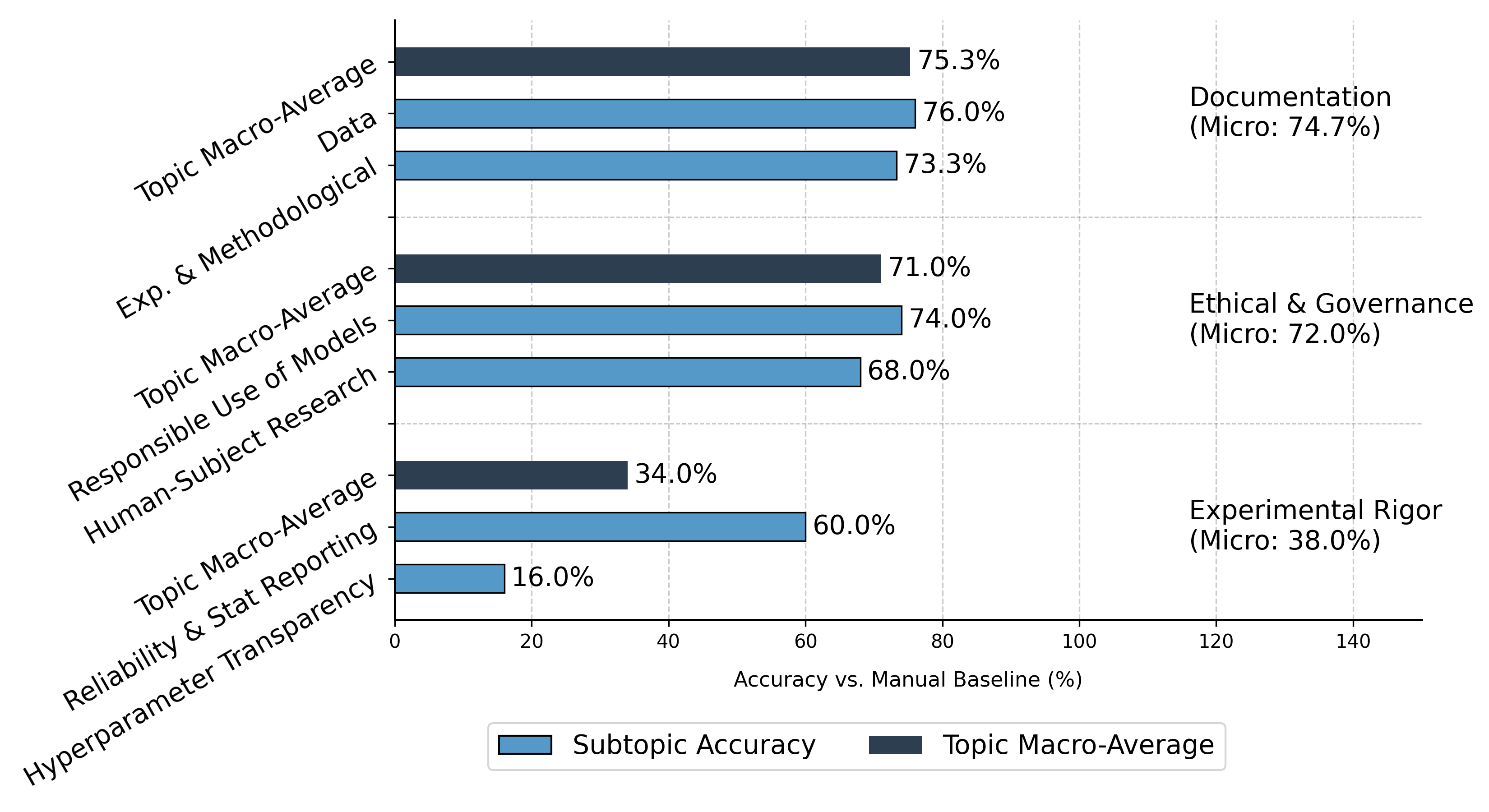}
    \end{subfigure}
    \caption{Accuracy benchmarked against the manual baseline, detailing both granular subtopic accuracy (blue) and the overall topic macro-average (dark slate).}
    \label{fig:subtopic_accuracy}
\end{figure}

Crucially, the intersection analysis highlights a massive performance variance depending on the evaluation topic. The agent performs well on structural verifications like \emph{Code and Environment} (84.6\%) and \emph{Data and Artifact Availability} (77.2\%), but performance plummeted on \emph{Experimental Rigor} (38.0\%). See \autoref{fig:subtopic_accuracy} for the full breakdown. Our review indicates that this discrepancy is bidirectional, a point we examine in \autoref{sec:discussion}.

Finally, qualitative semantic alignment with the human reports reflects this broader performance variance. We used an LLM-as-Judge to score the agent's semantic alignment against the human-authored ground truth (1--4). The agent achieved a mean alignment score of (2.99/4.00) for improvement suggestions, significantly higher than the (1.92/4.00) observed for evidence paths.  We examine this divergence in \autoref{sec:discussion}.

\begin{figure}[htbp]
    \centering
    \includegraphics[width=0.75\linewidth]{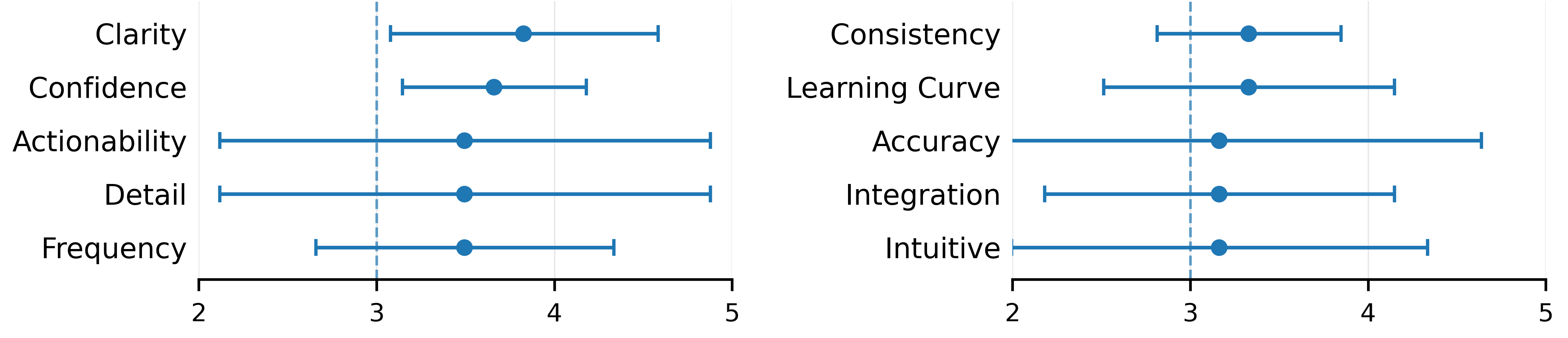}
    \caption{Descriptive participant ratings of the prototype ($n=7$). The five-point Likert scale ranges from 1 (strongly disagree) to 5 (strongly agree), with 3 as the neutral midpoint (vertical line). Points show item means and horizontal bars show $\pm 1$ standard deviation. Because this is a small pilot sample, the values summarize these participants only and should not be interpreted as population estimates.}
    \label{fig:survey_scores}
\end{figure}

\noindent\textbf{Pilot Survey.}
Across the seven survey participants (3 quantitative, 3 qualitative, and 1 mixed-method study authors), results indicate high usability and strong adoption potential. 
Within this small pilot sample, participants rated the tool’s clarity and integration positively, with subtopic understanding (Mean, standard deviation $M=3.83, SD=0.75$ Clarity) scoring highest and frequency-of-use intention ($M=3.50, SD=0.84$ Frequency) also high. 
Perceived usefulness was favorable, as participants felt confident using the tool to improve artifacts ($M=3.67, SD=0.52$ Confidence). 
However, qualitative consistency and report accuracy showed higher variance ($M=3.33, SD=0.52$ Consistency; $M=3.17, SD=1.47$ Accuracy; respectively), reflecting mixed performance across diverse methodology types. With only seven participants, these means and standard deviations can not be considered stable descriptive summaries and do not support statistical or population-level inference. The study-type counts likewise do not support subgroup comparisons. The qualitative responses provide initial explanations for the high variance.
Participants reported that the tool worked better for code-centric, quantitative artifacts than for complex qualitative or mixed-method material. 
\autoref{fig:survey_scores} lists the overall ratings and standard-deviations.

\noindent\textbf{Cost.}
Based on the Langfuse traces collected during development and evaluation, the per-paper cost of running the agent ranged from \$0.11 to \$2.23, with an average cost of \$0.71 per report.

\section{Discussion}
\label{sec:discussion}

The findings from this study demonstrate that agentic approaches can successfully operationalize broad open-science rules into automated checks, particularly for computational and code-centric research. 
The approach’s high inter-run consistency indicates that the inherent non-determinism of LLMs can be effectively constrained through explicit graph-based state management, structured tool use, and operationalized criteria. 
The proposed approach can handle high heterogeneity of replication packages reasonably well. 
Furthermore, the architecture introduces robust artifact ingestion that gracefully handles the structural heterogeneity and format diversity of replication packages. 
Unlike end-to-end models that generate opaque judgments, the agent strictly grounds its evaluations in extracted evidence. 
By mandating explicit file-path citations, the approach provides verifiable proof, allowing reviewers to independently audit its reasoning. 
This is supported by the high semantic consistency scores observed via the LLM-as-Judge evaluation for the agent's reasoning and suggestions.

The agent's bias toward strictness, reflected in its lower precision on non-compliant criteria, stems from a reliance on narrowly defined criteria that cause it to penalize artifacts organized in non-standard ways. For instance, when an author provides a unified \texttt{reproduce\_everything.sh} script rather than the one-to-one figure mapping the agent expects, the approach might flag it as a failure. The same rigidity drives the low agreement on \emph{Experimental Rigor}, where the agent's sub-condition checks diverge from the heuristics human reviewers apply. The challenge for the future is therefore to move beyond rigid, checklist-based verification and recognize when an alternative implementation is functionally equivalent to the expected standard. Likewise the low semantic agreement on evidence extraction reflects differing retrieval strategies. The agent's, more exhaustive, retrieval surfaces secondary, corroborating evidence that human reviewers bypass once a sufficient proof is found, paradoxically penalizing its alignment score.

Furthermore, the approach’s limited applicability to qualitative research remains a limitation. Qualitative research relies on context and researcher judgment, expressing rigor through traceability and logical consistency \cite{lincoln1985naturalistic}, rather than strict re-execution. Future work could explore the integration of structured qualitative reporting frameworks into machine-operationalizable representations. Fortunately, the approach’s modular knowledge base supports this extensibility. By separating the core execution logic from the underlying relational database, the approach can dynamically adapt to venue-specific checklists, publisher mandates, or emerging domain standards (e.g., guidelines for LLM studies \cite{baltes2025guidelinesempiricalstudiessoftware}) without requiring architecture modifications.

Rather than offering a simple pass-or-fail judgment, the approach diagnoses specific shortcomings and generates a to-do list for artifact improvement. However, user feedback regarding usability underscores a tension between this detailed transparency and cognitive load. As one participant noted regarding the Human-in-the-Loop (HITL) step: ``\textit{The planning step expecting human-in-the-loop guidance is overwhelming and it is very unclear to a new user, in how much detail they would need to understand and adjust the plan.}'' Future iterations of the frontend would benefit from adaptive interfaces that support progressive disclosure, allowing users to verify high-level decisions while exploring detailed reasoning traces only when explicitly needed. Ultimately, integrating such agentic evaluation approaches has the potential to reshape empirical software engineering research by shifting artifact evaluation from a retrospective, manual process to a scalable, proactive workflow. While confidentiality is crucial when reviewing unpublished work, our prototype relied on commercial models, which is appropriate at this early stage but of course unsuitable in the long run.

\section{Threats to Validity}
\label{sec:threats_validity}
We discuss the threats to validity of this study as outlined by Runeson and Höst \cite{Runeson2009}. \\

\textbf{Internal Validity.} Our prompting strategies may favor specific document structures or reporting styles, and model-specific traits, such as phrasing sensitivity, pretraining biases, and stochastic variations that we cannot mitigate within an emerging results report. The use of proprietary LLMs also risks implicit data leakage should the evaluated paper or public repository appear in their training datasets. We deem this risk as negligible.
Because the selected packages came from publications familiar to the authors, this familiarity may have simplified evidence discovery or influenced borderline judgments, thereby distorting measured agreement. We did not measure inter-rater reliability for the manual reference, so its own reliability remains unknown. The reference should therefore be understood as a fallible author-created comparison rather than ground truth.

\textbf{External Validity.} Our approach and underlying criteria are systematically biased toward computational, code-centric research, limiting generalizability to qualitative or theoretical software engineering paradigms. In our study, this expected bias stemming from the alignment of the open-science policies we drew on, caused the agent to inappropriately recommend restructuring a qualitative artifact into a code repository. Results also depend on the specific LLM versions used, and neither generalization across LLM versions nor comparison to other models was in scope of this work. 
The five purposively selected packages are too few to represent the diversity of replication packages in software engineering, and the seven recruited researchers are too few to represent the broader researcher population. Moreover, participants evaluated their own artifacts, so familiarity with and ownership of those artifacts may have influenced perceived usefulness, accuracy, and adoption intentions. We deem those risks acceptable for an emerging results report.

\textbf{Construct Validity.} Operationalizing reproducibility through a set of quality criteria may not fully capture its multifaceted nature in practice, a risk we accept as these criteria reflect the current state of practice. Employing an LLM-as-Judge adds a layer of approximation, relying on semantic, prompt-based comparison rather than deterministic verification, which we quantified through manual verification in \autoref{sec:results}.

\textbf{Reliability.} The non-deterministic nature of LLMs poses challenges for evaluation consistency. Although our prototype reached a high inter-run consistency, minor stochastic discrepancies persist across repeated executions. We mitigate this by triangulating automated judgements through multi-run aggregation and benchmarking against manual baseline assessments.

\section{Conclusion}
\label{sec:conclusion}
This study demonstrates that agentic workflows have the potential to operationalize high-level open-science policies into machine-verifiable tasks. Our prototype achieved high inter-run consistency and substantial agreement with manual baseline assessments. While the agent excels at offloading labor-intensive tasks (e.g., discovery and structural analysis) for quantitative, code-centric repositories, it exhibits a structural bias against qualitative research and unconventional artifact layouts. While its evidence-based traceability builds user trust, the interactive HITL phase can introduce a noticeable cognitive load, highlighting the need for a more streamlined user experience.

Our three contributions are a relational knowledge base of operationalized open-science criteria, a multi-agent prototype that autonomously retrieves, structures, and evaluates research artifacts, and a pilot survey on the usefulness of the approach. Together they take a step toward shifting artifact evaluation from a manual, resource consuming process, into a scalable, proactive workflow. Ultimately, we hope that in the future this approach can serve authors as a proactive ``replication package linter'' that flags quality issues prior to peer review, and supports artifact evaluation committees as a force multiplier that automates routine structural checks so reviewers can focus on deep semantic validation.

Future work will harden preprocessing against complex formats such as deeply nested archives and Jupyter Notebooks, integrate study-type detection for adaptive, context-aware criteria selection, apply ensemble methods such as majority voting to further stabilize consistency, and refine the frontend while adding support for open LLMs to broaden adoption. Finally, future work will have to evaluate the impact on resources in artefact evaluation tracks envisioned with our solution.

\paragraph*{AI Usage} 
We employed LLMs for textual editing, as well as diagram improvements.

\paragraph*{Data Availability}
The replication package supporting this study, including the scripts, raw repository data, agent configurations, survey instrument and data, is openly available on Zenodo at in our online material \cite{online_material} under a CC-BY 4.0 license. The replication package contains all materials necessary to reproduce our findings.

\bibliography{main}

\appendix

\end{document}